# User Empowerment in the Internet of Things

*Abstract*—This paper focuses on the characteristics of two big triggers that facilitated wide user adoption of the Internet: Web 2.0[1] and online social networks. We detect brakes for reproduction of these events in Internet of things. To support our hypothesis we first compare the difference between the ways of use of the Internet with the future scenarios of Internet of things. We detect barriers that could slow down apparition of this kind of social events during user adoption of Internet of Things and we propose a conceptual framework to solve these problems.

## I. INTRODUCTION

Apparition of Web 2.0 changed the way people use the internet. It allowed users to participate actively in creation of digital content, digital information sharing and collaboration over the Internet. Newly created web applications became more user-centered and enriched the user experience. Often used as web services the web applications allowed the users to compose their own set of services and personalize their use of the Web.

We observe recent propagation of online social networks whose creation was facilitated with the development of the Web 2.0. The web users increased collaborative creation and sharing of digital content and services that is more similar with their social interactions. The access to the user created digital content and services that was often treated on the web as *public or private* is evolving in the access for *friends* or *friends of the friends*. With the wide adoption of Web 2.0 and online social networks user's requirements for privacy are lowered[2] for the benefits of social life and interactions trough the online social networks. Not only are the end users involved in this evolution but also commercial actors. They are also evolving from official communication with their customers through the static web sites to more friendly relationships trough the social networks.

Better user experience and social shape of the Web increased significantly the number of users and changed the way people interact on the Web.

In this paper we advocate for the user control of his connected objects, services provided trough that objects and his privacy in emerging Internet of things (IoT). We first compare the important characteristics that triggered the proliferation of the Web 2.0 with the characteristics of emerging IoT. This analysis lead as to a definition of our conceptual framework intended to define the work boundaries. We than propose a working scenario from which we identify problems and issues that could prevent user adoption of IoT ecosystem. Based on detected problems and issues we propose the future works in order to make IoT more user-friendly and more social. We believe that the integration of IoT in user's social life can lead to wide IoT adoption.

## II. WHY INTERNET OF THINGS IS DIFFERENT

Although the research in smart objects roots back to Marc Weiser's vision of ubiquitous computing[3] valuable research results in wearable computing, ambient intelligence and smart environments are not yet widely merged into our everyday lives. The question of interoperability between smart objects, applications and services still remain open. European commission is making significant efforts in governance and standardization for the network of connected objects, protection of private life and security of connected objects[4]. This initiative together with industrial efforts for standardization[5] will bring the seamless integration of smart objects and industrial products into the Internet.

The importance that the IoT follows the experience of Web evolution has already been emphasized[6]. Due to ubiquitous nature of IoT and increased number of devices and objects surrounding user we focus here our attention on user's control, user's social life and user's privacy. With this in mind we analyze the difference between the Web 2.0 and the IoT.

### A. User's control

According to Marc Weiser's vision of ubiquitous computing the mature technologies are disappearing from the sight, users only know and see the final results provided by technology[3]. As the IoT and ubiquitous computing are still not mature technologies they are still didn't reach this stage of being invisible per users. Users must know how to use their connected objects, how to consume services provided by these objects and how to personalize all of it. We believe that the rules from human machine interactions of immediate feedback and keeping user aware of all interactions used in Web design are not adapted for IoT due to ubiquitous presence of connected objects. Prompting the users in every situation or every place about the interactions between his objects and services would be too disturbing and overwhelm people in their everyday lives. Research directions in artificial intelligence and expert systems tempt to solve this problem by making objects and environments able to take decisions for the users. Although research result from these directions help making connected objects more manageable they are not necessarily making collaboration between people more social.

We believe that the user control over connected objects is a crucial element for adoption of the IoT. This control should not be considered only as a single user control over his connected objects, it should allow users to share the information about objects and the services provided by these connected objects. As the Web 2.0 gave the control to the end users allowing them to share the digital content and interact trough the web services, the IoT should allow users to choose how they share the information about connected objects and to choose derived services.

*B. User's social and virtual Life*

On the Web, users are creating their digital identities[7] and interact between virtual communities and online social networks. They create digital identities willingly for virtual places they are visiting. On online social networks users choose when to connect and share only the selected digital content. The action of creating a digital identity and submitting the digital content is always initiated by users.

In the IoT the user's real life is tightly connected to his virtual life. When the personal objects are mapped in the network the action of submitting the digital information by the objects can be out of user's control. With the recent proliferation of online social networks we observe that people are willing to connect with real friends and share the digital content on virtual places. We believe that the users should be able to share the digital information about their objects with their online social relationships. The sharing of digital information between social relationships like friends would help the adoption of the IoT.

*C. User's privacy*

The question of privacy is still an open question in the IoT as well as in ubiquitous computing. The problem arises from the questions of usage and possession of digital information. As stated in the communication of European parliament for IoT, it is likely that the uptake of IoT will affect the way we understand the privacy[4]. Anyhow, with the wide use of online social networks our privacy is already evolving especially between younger generations.

We believe that the privacy should not be considered in the IoT as traditional right to access the digital information, it should be considered as a social sharing. The IoT should embed the information from the smart and industrial objects into the networks of friends and bring more friendly services.

### III. REFERENCE FRAMEWORK

We propose a conceptual framework for user adoption of IoT trough the social interactions. We believe that the social benefits perceived by end users can help emerging IoT to be adopted and to reach the critical mass.

Before we introduce the problems from our scenario, define our reference framework:

We use the term **"People"** in order to describe the real persons, users in emerging IoT. Virtual identities are not directly considered by this term.

We define **"Position"** as geographical point on earth defined with the latitude, longitude and altitude[8].

The term **"Space"** is a three dimensional reality surrounding the user[9].

The term **"Virtual Space"** is a virtual space accessible user.

The **"Place"** is user's definition of the space and an administrative space.

As an example we can say that the restaurant from our scenario has the position of 46012'00''N and 6009'00''E, the interior of the restaurant is the administrative space surrounding Alice during working lunch which could also be augmented virtually and finally the restaurant is defined by Alice as her working place.

We use the term **"Things"** in order to describe interchangeably user's electronic devices and industrial products tagged by Radio Frequency Identifiers (RFID) connected directly or indirectly to the Internet.

### IV. SCENARIO

In previous chapters we briefly compared the characteristics of two events that made the Web more usable and more social. From the detected characteristics we define a working scenario that describes a usual day and the social life of an empowered user in IoT. Based on this scenario we detect challenges to empower the users and merge the new IoT technologies in their social life.

*24 Hours of Alice in the Internet of Things*

Alice is having the breakfast in her kitchen; her kitchen table displays the possible meals from available food at home and suggests the recipes for dinner when she gets back. Alice would like this time something else and she chooses other recipes. Shortly after she confirms the meal she chose for dinner. The table transmits the food order automatically to Alice's favorite shopping place and adds a new reminder to her smart phone to pick up the purchase on her way home. When she exits the house the house reduces the heating, closes all windows and activates the alarm. While she is opening the door of the car, the car recognizes her; it adjusts the seats, puts Alice's favorite radio station and plans the best route to work avoiding the traffic jams.

At the work, when she enters the building Alice's computer boots and when she arrives in the office it displays her daily planning. Her boss scheduled a meeting in the afternoon but she also received multiple meeting propositions for lunch break. She accepted one and her agenda confirmed one more lunch reservation in the restaurant. During the lunch time the working group is discussing about different topics. They put their smart phones at the restaurant's smart table where they lunch and they can exchange the visit cards, next appointments and other useful digital information just by drag and drop of digital content between the smart table and their smart phones.

After the working after noon Alice decides to buy a new smart phone and she passes by a shop in the town center. She bought a smart phone and took it in the box together with a new SIM card. While she was still in the town her smart phone displayed the reminder with her shopping list for diner.

The smart phone estimates that she cannot arrive to her favorite shopping place before it closes and suggests a near shop that has the same articles. Alice thinks that it is a good suggestion and does the shopping in the suggested shop.

After the shopping Alice drives home and when she arrives the home temperature is at the level she likes. While entering the house Alice's smart home assistant displays digital bills that arrived during day on the entry wall. Alice just gives an order with the natural voice that she will treat that later. First she wants to cook than she will take care about the bills and she will setup her new smart phone. As Alice likes cooking in her spare time she also likes to share the cooking tips with her friends. When she chooses the evening recipe on the kitchen's smart table she also chooses "social option". Her kitchen is connected to a social network and her friends can see what she is preparing for dinner. After a few minutes her friend Lara is online and her video call is projected on kitchen's wall. They exchange a few cooking tips but they also decide to go out for a drink and meet also other people. They choose the time and the bar where they will meet from the map displayed on the kitchen's wall. After diner Alice goes to her living room to pay the bills and to setup her new smart phone. From her smart table in living room she picks up the mailbox option, reads the bills and in order to pay them moves them to the digital representation of her credit card. After the bills she unpacks her new smart phone. She puts the SIM card and she starts device. The smart phone boots, it finds the operator network but it finds multiple home networks from Alice and her neighbors. Alice chooses her home network and provides her password. The smart phone detects her home devices and it suggests synchronizing of her accounts, contacts, emails and bookmarks from her personal computer. It also suggests downloading the installed applications that Alice had on her previous smart phone. After the synchronization the smart phone also suggests a choice of new applications. She installs one that is called "keep me offline". Quite satisfied with this easy smart phone configuration Alice puts it on the charger and prepares to go out.

When Alice enters the car she pronounces loudly her destination address and the car's GPS replies her that it's ready to guide her and that her Friend Lara will also arrive on time. In the bar Alice meets Lara and they choose a free smart table for two people. While chatting in the bar Alice and Lara put their smart phones on the table surface. As the restaurant's smart table also knows that they like fruit juices and that they came by car it suggests them a choice of natural and non alcoholic drinks. After ordering their drinks from the table surface they can display the maps with the position and activities of their friends. The maps on the smart table surface also have a small application called "party planer" and they decide to try it out. This application suggests them an approximate central position among their friends which are present at a rock concert and let Alice and Lara to choose time and to notify their friends to meet there. After the drink they pay their bill on the smart table by dragging the digital bills in direction of their smart phones and they are ready to go to the concert.

Late in the night, after the party, Alice took her car and she is on the way home. As her car detected that it is soon time for changing the oil it displayed the agenda with suggested free slots for its service. Alice was pretty tired to decide this meeting now and she didn't want to be disturbed any more by non urgent messages so late in the night. She took her smart phone and started the "keep me offline" application. The agenda for car service immediately disappeared from the car's screen. Instead the car will display a small red button in the corner of the screen only if its state becomes critical. Alice's friends cannot know any more where she is, her kitchen at home will not report that she had a glass of milk before going to sleep but she knows that her house will be warm when she arrives.

## V. ISSUES AND PROBLEMS

From the above scenario we can detect several problems to be considered during design of IoT but for our conceptual framework we group these problems according three points of view: People, Places and Things. The goal of this problem classification is to focus on user empowerment for everyday usage and social life including connected objects at the different places.

### A. Objects and information ownership

The first problem category that we detect from our scenario is coming from the object's ownership. In our scenario Alice uses her own objects, the objects that belong to her employer and the public objects. Regarding Alice's privacy not all objects should share all available information about her. Regarding usage and control over the objects she will have more time and privileges to personalize her own objects than the objects that are not in her possession.

From our scenario some objects are not completely in Alice's possession. Some objects belong to Alice's employer and some to public places like restaurants. Personal objects should be able to recognize their owners and to adapt to them. Other objects should adapt them self to public users without complete information about the user.

In our working scenario Alice interacts with her friends, working colleagues and people in the store when she is buying a new smart phone. Trough everyday objects she can also interact implicitly with many other people. We detect two main problems under relationship between people trough everyday objects: how do people control the access to their personal information and are they always able to personalize everyday objects. From our scenario Alice shouldn't be in doubt if her contacts will be recorded and distributed by the smart table in the restaurant during her work lunch. Alice should trust the restaurant that provides the object. From the point of view of the restaurant the smart table could be rented from another firm in order to provide an extra service. In this case we believe that the restaurant should not use the model of license agreement which is often used to rule the relationships between users and service providers on the Web. We believe that this model is not adapted for IoT ecosystem. Due to ubiquitous objects presence users cannot read license agreement in every situation.

Instead we propose a logo based types of agreements that group the terms of use in easily understandable representations[10].

One could be taken into a premature conclusion that we could classify objects in function of their place. For example in our scenario Alice's objects at work could be the objects in possession of her employer and meant to provide her assistance for work. But then we have a situation where Alice is in a restaurant for a working lunch and her working group uses the restaurant's smart table for work. We also have the situation where Alice shares her cooking recipes from her personal kitchen table. From these examples we can exclude the definition of object's ownership and behavior in function of their physical places. We rather define places as the physical places enriched with personalized meaning. From our working scenario Alice should be able to give the meaning of "*working place*" to a public restaurant when she has a work meeting and she should also be able to give meaning of "*public place*" to her kitchen but only when she cooks. The situation is even more complex if we add the notion of virtual places. Some objects or their functions can be used from distant physical places. From this assumption we detect the second problem: how Alice's can easily create a working place in the public restaurant if she doesn't give all of her personal information to the surrounding objects and she doesn't have time to personalize these objects manually?

We believe that the user can have a set of predefined privacy policies that could determine objects behavior between personal and external objects.

*B. Personalization*

In order to accomplish the above scenario the IoT ecosystem should allow people to be in control of everyday objects in effortless manner. The works on human machine interactions, ubiquitous user models[11] and design of everyday objects[12] are bringing continuous improvements of objects usage. With continuous development in these domains people have significant help to use numerous objects. Still, user input, control over object interactions, personalization of an object or federated objects is a vital element of technology adoption. We believe that people should be in permanent control over the objects they use during everyday life.

In IoT ecosystem objects should use universal plug and play protocols in order to be seamlessly integrated in user environments. They should also be able to distinguish their owners from their occasional users. Object should use universal protocols in order to establish ubiquitous communication but they should also distinguish what user information they share between personal and shared objects.

Objects should be able to adapt to the user's needs according the context and place. From our example Alice's smart phone should perform different tasks at work than at home. We emphasize here that the place should be user defined and not only the geographical position of the objects.

Our scenario describes Alice's kitchen not only as a physical place but also as a virtual place. The kitchen's table is showing the weather forecast and the kitchen's wall can communicate with Alice's friends. From this situation we observe that the everyday objects even situated at the different positions can communicate to each other thus creating a virtual space for Alice. The objects near Alice can be context aware but despite the context awareness of everyday objects Alice can decide at any moment to change the meaning of her surrounding space. She can decide to use her kitchen for work and in this situation the contextual information about Alice being in her kitchen is not sufficient for her surrounding objects. Alice can initiate an action that can augment her context for ex. She can push the button "*working space*" on her kitchen's table. In that case her home objects should discover and virtually integrate the objects from her office into her virtual space.

## VI. CONCLUSION AND FUTURE WORK

For the future work we propose a user-centered system design of a social platform. The goal is to solve previously detected problems in order to empower users for their social life in emerging IoT. We propose an implementation of the framework based on People, Places and Things that will allow users to manage and share the digital information about the objects and services. As for the introduction we give the following schema of user's social interactions.

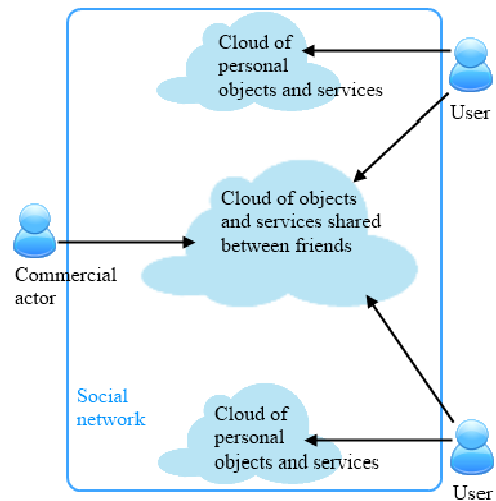